\documentstyle[aps,tighten,epsf]{revtex}

\begin{document}

\draft

\title{Bremsstrahlung in intermediate-energy nucleon reactions within an
effective one-boson exchange model}

\author{
{\sc V.V. Shklyar$^{a,b}$,
B. K\"ampfer$^{c}$,
B.L. Reznik$^b$,
A.I. Titov$^{a,b}$\\[6mm]}}

\address{
$^a$Bogoliubov Theoretical Laboratory, Joint Institute for Nuclear Research,\\
141980 Dubna, Russia\\[1mm]
$^b$Far-Eastern State University, Sukhanova str. 8,\\
Vladivostok 690900, Russia\\[1mm]
$^c$Research Center Rossendorf Inc.,
Institute for Nuclear and Hadron Physics,\\
PF 510119, 01314 Dresden, Germany
}

\maketitle

\begin{abstract}
Within a covariant effective one-boson exchange model for the
$T$ matrix of $NN$ interactions we present detailed calculations of
bremsstrahlung cross sections
for proton - proton and proton - neutron reactions
at beam energies in the 1 GeV region. Besides pure bremsstrahlung
processes we consider photons from $\Delta$
decays and contributions
from the $\eta \to \gamma \gamma$ process.
At beam energies above 700 MeV
the $\Delta$ decay channel dominates the spectra at large photon energies,
where the interference between non-resonance processes and the
$\Delta$ decay channel becomes also important.
Low energy photons stem from pure bremsstrahlung processes.
The available experimental data at 730 MeV
beam energy is well described.
We extrapolate the model down to 280 MeV, where more detailed experimental
data exist, and find agreement with angular distributions.
\end{abstract}
\pacs{\\ \vskip 1cm
      {\it Key Words\/}: bremsstrahlung, effective one-boson exchange model\\[0.7cm]
      {\it PACS\/}: 13.75.Cs, 25.20.-x, 25.40.-h}


\newpage

\section{Introduction}

The series of experiments with proton
kinetic energies $T_{\rm kin} =$ 280 - 2500 MeV
at the new cooler synchrotron (COSY) in J\"ulich
promises a wealth of precision data of various hadronic reactions.
Among the particle production experiments a systematic study of
bremsstrahlung processes is envisaged \cite{LoI}, and the first data
with the time-of-flight detector system TOF are analyzed \cite{Schulke}.
The interest in bremsstrahlung is as usual for electromagnetic processes
in strongly interacting systems: real or virtual photons serve as
clean probes of the strong interaction.

In the energy region up to the pion production threshold at
beam energy of $T_{\rm kin} \simeq$ 280 MeV the bremsstrahlung can test the
half-off-shell part of the nucleon - nucleon ($NN$) potential. One motivation
for such experiments was therefore to constrain the various potential models
of the low-energy $NN$ interaction. Recent studies
\cite{deJong,Herrmann,Jetter}
however show that several bremsstrahlung observables are not
too sensitive to different parametrizations of the $NN$ potential.
New precision data might allow for some progress, since the available data set
in this energy region \cite{Kitching,Michaelian}, as
noted in Refs.~\cite{Herrmann,Jetter,Scholten,Herrm2}, does not point to a
preferable $NN$ potential.
Indeed, various models can account for the data \cite{Jetter,Scholten,Eden},
even if there is some uncertainty concerning the absolute
normalization of the cross section \cite{Michaelian}. Some new
possibilities to distinguish among various potential approaches
in bremsstrahlung reactions are proposed recently in Ref.\cite{Neudachin}.

At higher beam energy there is only one data set available,
namely at $T_{\rm kin} =$ 730 MeV \cite{Nefkens} where the energy spectrum
of the outgoing photons in limited phase space regions has been measured.
This data has been analyzed early \cite{Tiator} and also recently
in Ref.~\cite{Jetter} within covariant models
based on $\pi$ and $\rho$ exchanges. At COSY the measurements
can be extended in a broad beam energy interval
up to 2.5 GeV and deliver angular distributions of photons
and recoil nucleons. Theoretically, one must utilize appropriate
covariant strong interaction models in this energy region.

In particular the covariant one-boson exchange (OBE) models represent
a powerful tool for describing hadron reactions in the GeV region.
Such models have been employed for analyzing
\cite{dileptons,dileptons1,dileptons2,Mosel1} the
virtual photon (i.e., di-electron) production \cite{DLS}.
Recently, also the OBE description of pion production finds
renewed interest \cite{Moselpi}.
Therefore, it seems
worthwhile to apply the same model to the real photon production processes
to arrive at a coherent picture of electromagnetic processes in
strong interactions of nucleons.
Such a task is particularly tempting as there are controversial explanations
of peculiarities of the virtual photon production mechanism at
$T_{\rm kin} =$
1 - 1.5 GeV \cite{dileptons,dileptons1,Mosel1}: Either subtle interferences
between non-resonance and resonance (including the $\Delta$) channels
or a strong contribution of $\eta$ Dalitz decays can explain the non-trivial
beam energy dependence of
cross sections. Measurements of refined observables \cite{Bratkovskaya}
might disentangle the various di-electron sources. On the other hand an
independent test of real photon production in the
mentioned energy region may be important for studying this problem as well.

Here we present results of detailed calculations of the real photon
bremsstrahlung within a relativistic and gauge-invariant effective
OBE model, which continues and extends earlier investigations
\cite{dileptons1,Kamal,Mosel,PLB}
for the reaction $NN \to NN \gamma$.
We go beyond our previous study \cite{PLB} by including
the internal radiative meson conversions,
the anomalous magnetic moment of the nucleons
and the tensor $\rho NN$ coupling, and
a wider beam energy interval.
Conceptually, our approach is close to the  model of Ref.~\cite{Mosel1}.
The main difference rests in a different
goal of our study: We here consider angular and energy distributions
of real photons instead of inclusive invariant mass
distributions of di-electrons as in Ref.~\cite{Mosel1}.
Additionally we employ a slightly different
prescription for the $\Delta$ excitation vertex and effective parameters
in the two-body elastic scattering $T$ matrix.
In relation to Ref.~\cite{Jetter} where only
$\pi,\rho$ exchange has been considered, we include the exchange
of $\sigma$ and $\omega$ mesons as well and we
take into account
radiation from internal lines with vector mesons $V = \rho, \omega$
in $V\pi\gamma$ vertices.

Our paper is organized as follows.
In Section 2, we formulate our effective OBE model approach for
non-resonance and resonance contributions
to the exclusive reaction $NN\to NN\gamma$.
Concerning the resonance contributions we restrict
our consideration to the implementation of the $\Delta$ excitation and decay
channel. In Section 3 we analyze contributions of different channels
in the 1 GeV region and present a comparison with the available experimental
data for $pp \to pp\gamma$ at beam energy of 730 MeV.
In Section 4 we discuss the contribution of the $\eta \to \gamma\gamma$
decay channel to the $NN\to X\gamma$ reaction and compare it
with the strongly interfering $\Delta$ decay process and non-resonance
bremsstrahlung channels. In Section 5 we compare our model with data
at 280 MeV and discuss whether the effective two-body $T$ matrix model
may be applied here. The summary is given in Section 6.
In the present paper we restrict ourselves to unpolarized observables.
The analysis of polarization effects in $NN$ bremsstrahlung
will be subject of a forthcoming paper.


\section{$NN\to NN\gamma$ bremsstrahlung}

The differential cross section of
the exclusive reaction $N_1 N_2 \to N_1' N_2' \gamma$ reads
in a Fermion - Boson symmetric convention
\begin{eqnarray}
d \sigma( \sqrt{s})=
\frac{1}{2(2\pi)^5\sqrt{s(s - 4 M_N^2)}} \,
| \overline{{\cal M}} |^2  \,
\frac{d {\bf p}'_1}{2E_1'} \,
\frac{d {\bf p}_2'}{2E_2'} \,
\frac{d {\bf k}}{2\omega}  \,
\delta^{(4)} (p_1+ p_2- p'_1-p'_2-k),
\label{eq1}
\end{eqnarray}
where $k=(\omega, {\bf k})$ denotes the four vector of the outgoing photon,
$p_{1,2}=(E_{1,2},{\bf p}_{1,2})$ and $p'_{1,2}=(E'_{1,2},{\bf p'}_{1,2})$
are initial and final nucleon four-momenta, respectively.
$\sqrt{s}$ is the total available energy, and
the nucleon mass is denoted by $M_N$.

We do not consider here polarization observables,
therefore the matrix element
$|\overline{{\cal M}|^2}$ is spin averaged in the entrance
channel and spin summed in the exit channel.
By partially integrating over energies and directions of the outgoing
particles one arrives at experimentally relevant quantities,
e.g.,
$d\sigma/d \omega d \Omega_\gamma d \Omega_{p'}$
where $\omega$ stands for the photon energy in the center-of-mass system
(cms), and $\Omega_\gamma$ and $\Omega_{p'}$ denote the cms angles
of outgoing particles.
This differential cross section can be written as
\begin{eqnarray}
\frac{\omega d^5 \sigma}{d \omega \,d\Omega_\gamma d\Omega_{p'}}
=
\frac{\omega^2}{8 \, (2\pi)^5 \sqrt{s \, (s - 4M_N^2)} } \,
| \overline{{\cal M}} |^2 \, D(p'),
\label{eq2}
\end{eqnarray}
where
\begin{eqnarray}
D(p') = \frac{p'^2}{ |A p' + C E'|},
\label{eq3}
\end{eqnarray}
with
\begin{eqnarray}
&&E' = \frac{AB - C \sqrt{B^2 -M^2_N (A^2 - C^2)}}{A^2 - C^2},\nonumber\\
&&A = 2 (\sqrt{s} - \omega),\,\,\,
B = s - 2 \omega \sqrt{s},\,\,\,
C = 2 \omega \cos \Theta_{\bf p'k}.
\nonumber
\end{eqnarray}
The 3-momentum of the outgoing proton in the cms
is $ {\bf p}'$ with $p'^2 = {\bf p}'^2 = E'^2 - M^2_N$.

We perform diagrammatic calculations of bremsstrahlung processes on the
basis of the effective OBE model, where four exchanged mesons $M =$
$\pi^{\pm,0}, \rho^{\pm,0}, \omega, \sigma$
are used for the description of the two-body $pp$ and $pn$ $T$ matrices.
The resulting Feynman diagrams on the tree level
which contribute to the amplitude
${\cal M}$ are displayed in Fig.~1. Let us first consider
the $pp$ reaction. The amplitude
\begin{equation}
^{pp}{\cal M} =
\sum_{M=\pi^0,\sigma,\omega,\rho^0}
{^{pp} \hat {\cal M}^{(1)}_M}
+
\sum_{MM'=
V^0, \pi^0}
{^{pp} \hat {\cal M}^{(2)}_{MM'}}
+
\sum_{M=\pi^0,\rho^0}
{^{pp} \hat {\cal M}^{(3)}_M}
\label{eq4}
\end{equation}
is a coherent sum of
(i) pure bremsstrahlung diagrams
$\hat{^{pp}{\cal M}^{(1)}_M}$
with pre-emission and post-emission of a photon with momentum $k$ as shown
in Figs.~1a - d, where the intermediate nucleon is an off-shell proton,
(ii) the contributions from internal radiative meson conversions
$\hat{^{pp}{\cal M}^{(2)}_{MM'}}$ according to the processes
$V^0 \to \pi^0 \gamma$ (where $V^0$=$\rho^0,\omega$)
displayed in Fig.~1f, and
(iii) resonance contributions
$\hat{^{pp}{\cal M}^{(3)}_M}$
from excitation and decay of the $\Delta^+$(1232) resonance
as depicted in Figs.~1a - d.
In (ii) we neglect possible $V\to\eta\gamma$ transitions
because the corresponding
coupling constants are expected to be negligible small \cite{PDG}
and we envisage a complete calculation with $\pi$, $\rho$, $\omega$, $\sigma$
exchange.
Due to identical particles in the initial and exit channel the corresponding
exchange diagrams must be included, i.e.,
$${^{pp} \hat {\cal M}^{(n)}_{\cdots}}
=
{^{pp}{\cal M}^{(n)}_{\cdots}} (p_1,p_2;p_1',p_2') -
{^{pp}{\cal M}^{(n)}_{\cdots}} (p_1,p_2;p_2',p_1'),
\quad n = 1, \cdots, 3.$$
This results in a total of 48 diagrams, i.e., 32 pure bremsstrahlung
contributions, 8 graphs with $\Delta$ excitation, and 8 diagrams
with internal meson conversion.

The matrix element for the reaction $pn \to p n \gamma$ reads accordingly
\begin{eqnarray}
^{pn}{\cal M} & = &
\sum_{M=\pi^0,\sigma,\omega,\rho^0}
{^{pn}{\cal M}^{(1)}_M}
+
\sum_{M=\pi^\pm,\rho^\pm}
{^{pn}{\cal M}^{(2)}_M}
+
\sum_{M=\rho^\pm,\pi^\pm}
{^{pn}{\cal M}^{(3)}_{M}} \nonumber \\
& &
+
\sum_{MM' = V,\pi}
{^{pn}{\cal M}^{(4)}_{MM'}}
+
\sum_{M=\pi^0,\rho^0}
{^{pn}{\cal M}^{(5)}_M}
+
\sum_{M=\pi^\pm,\rho^\pm}
{^{pn}{\cal M}^{(6)}_M}
\label{eq5}
\end{eqnarray}
and contains the pure bremsstrahlung processes with exchange of neutral
and charged mesons ${^{pn}{\cal M}^{(1,2)}_M}$
(Figs.~1a - d),
radiation from charged internal mesons ${^{pn}{\cal M}^{(3)}_{M}}$
(Fig.~1e),
internal radiative meson conversion ${^{pn}{\cal M}^{(4)}_{MM'}}$
(Fig.~1f),
and resonance contributions from the $\Delta^{+,0}$ decays
${^{pn}{\cal M}^{(5,6)}_M}$ (Figs.~1a - d).
This yields 38 diagrams in total, i.e., 24 pure bremsstrahlung contributions,
8 graphs with $\Delta$ excitation, and 8 diagrams with internal meson
conversion.
Note that this exhausts the full set of one-boson exchange diagrams
for the exclusive one-photon bremsstrahlung reaction with
four exchanged bosons.
The contributions from the inclusive processes depicted
in Figs.~1g and h in $pp$ and $pn$ reactions will be separately discussed
in Section~3.

\subsection{Non-resonance pure bremsstrahlung}

To evaluate the mentioned matrix elements
we employ the covariant perturbation
theory in first order to nucleon - nucleon interactions.
The nucleon - meson interaction Lagrangian reads
in obvious standard notation
\begin{eqnarray}
{\cal L}_{int} =
-i g_\pi \bar\psi \gamma_5 \bbox{\tau} \psi \bbox{\pi} +
g_\rho \bar\psi \bbox{\tau}
[\gamma_\mu  \bbox{\rho}^\mu + \frac{\kappa_\rho}{2 M_N} \sigma^{\mu \nu}
\partial_\nu \bbox{\rho}_\mu] \psi +
g_\omega \bar\psi \gamma_\mu \psi \omega^\mu +
g_\sigma \bar\psi \psi \sigma,
\label{eq6}
\end{eqnarray}
where $\psi$ represents the nucleon bispinor, and
$\bbox{\pi}$, $\bbox{\rho}_\mu$, $\omega_\mu$, $\sigma$ stand for the
respective meson fields; $\mu, \nu \cdots$ are used as Lorentz indices,
and $\bbox{\tau}$ is the Pauli matrix.
Symbols in boldface denote here isovectors.
Note that we employ the
pseudo-scalar $\pi NN$ coupling. It is known that the pseudo-vector
and pseudo-scalar couplings give analogous results
on the tree amplitude level \cite{Mosel}.
The strength parameter of the $\rho$ tensor current is $\kappa_\rho =$ 6
\cite{Jetter,Mosel1}; for the $\omega$ meson such a term is
negligibly small.

The strong $NNM$ vertices are dressed by monopole form factors:
either for each vertex
\begin{eqnarray}
F_{NNM} (q^2) = \frac{\Lambda_M^2 - m_M^2}{\Lambda_M^2 - q^2},
\label{eq7}
\end{eqnarray}
when the internal meson line carries the momentum $q$,
or for both vertices
\begin{eqnarray}
F_{NNM}(q_1^2)F_{NNM'} (q_2^2)
\left(1 +
\frac{m_M^2 - q_1^2}{\Lambda_M^2 -q_2^2} +
\frac{m_M^2 - q_2^2}{\Lambda_M^2 -q_1^2} \right),
\label{eq8}
\end{eqnarray}
when the internal line radiates a photon with momentum
$k = \pm (q_1 - q_2)$; the $MM'\gamma$ vertex is then
\begin{equation}
\Gamma_{MM\gamma}^\mu = -i e (q_1 + q_2)^\mu,
\label{eq9}
\end{equation}
with $e$ as electromagnetic
coupling.
If the photon is emitted from a nucleon line the $NN\gamma$ vertex  function
is as usual
\begin{eqnarray}
\Gamma_{\gamma NN}^\mu =
-i e\left[\gamma^\mu F_1+i\frac{\kappa_N}{2M_N}\sigma^{\mu\nu} k_\nu
F_2 \right],
\label{eq10}
\end{eqnarray}
where $\kappa_N$ is the anomalous magnetic nucleon moment
($\kappa_N =$ 1.79 for protons and -1.91 for neutrons)
in terms of the nuclear magneton,
and $F_{1,2} =$ 1 for protons and $F_1 =$ 0, $F_2 = $ 1 for neutrons.
The form factors are constructed in such a way
that they preserve gauge invariance \cite{Mosel1,Haglin,gauge}.

For the elastic $NN$ scattering, the strong $NNM$ vertex form factor (7)
depends only on the momentum transfer squared $q^2$.
For the half-off-shell $T$ matrix, which appears in the
bremsstrahlung processes, the effective vertex functions, in general,
must depend on an additional invariant variable. The momentum
squared $p^2$ of the off-shell nucleon may be chosen as such a variable.
A procedure of including the off-shell dependence into both the
electromagnetic form factors $F_{1,2}$
and two-body $T$ matrices is discussed in
Ref.~\cite{Mosel1}, and it is argued that the off-shell correction
is expected to be not so large.
In order to avoid in
the present consideration poorly constrained additional parameters,
we assume that the above adjusted on-shell $T$ matrix parameters
are applicable also for the half-off-shell amplitudes.

In our calculations we use energy dependent parametrizations
of the elastic $pp$ and $pn$ scattering,
where the effective coupling constants $g_M$
change with beam energy according to
$g_M(s) = g_M^0 \exp(- \beta_M \sqrt{s - 4 M_N^2})$
\cite{scaling}.
The meson masses $m_M$, coupling strengths $g_M$, and cut-off parameters
$\Lambda_M$
are adjusted by standard fits to the known experimental data
and are listed in Table~1 together with the energy parameter $\beta_M$.
In Fig.~2 we show the results of our calculations for the
elastic NN scattering within the effective OBE model in comparison
with experimental data \cite{cross_sections,elsc}
at two relevant beam energies.
It should be emphasized that the notion of the OBE model is to be understood
in the sense that a process is determined by the set of tree level
Feynman diagrams with given exchange bosons.

Together with the conventional photon radiation from the charged
exchange meson lines we include the radiation from internal
radiative decays as described above. The corresponding decay vertex reads
\begin{eqnarray} \label{Vpg}
\Gamma_{V\pi\gamma}^{\nu\beta}=-ig_{V\pi\gamma}\epsilon^{\mu\nu\alpha\beta}
k_\alpha q_\mu,
\label{eq11}
\end{eqnarray}
where $V$ denotes the vector mesons $\omega$ and $\rho$,
and $q$ and $k$ are the pion and photon
momenta; $\epsilon^{\mu\nu\alpha\beta}$ is the Levi-Civita symbol.
The coupling constants $g_{V\pi\gamma}$
are adjusted to the experimental $V \to \pi\gamma$
decay widths \cite{PDG}. Their numerical values are
\begin{eqnarray}
\frac{g^2_{\omega\pi\gamma}}{4\pi} = 3.94 \cdot 10^{-2}\,
{\mbox{GeV}}^{-2},\,\,
\frac{g^2_{\rho^0\pi^0\gamma}}{4\pi} = 0.692 \cdot 10^{-2}\,
{\mbox{GeV}}^{-2},\,\,
\frac{g^2_{\rho^\pm \pi^\pm\gamma}}{4\pi} = 0.396 \cdot 10^{-2}\,
{\mbox{GeV}}^{-2}.
\label{eq12}
\end{eqnarray}
The choice of a positive sign for the $\omega\pi\gamma$ amplitude
and a negative sign for $\rho\pi\gamma$ is consistent with
the  data on pion photoproduction \cite{pion-photo}.
The $NNV$ vertices in $V\pi\gamma$ decay diagrams
are the same as discussed above for the conventional
OBE amplitudes with the same form factors and
energy dependence of the coupling constants.


\subsection{$\Delta$ excitation in $NN$ bremsstrahlung}

A second subgroup of contributions
to $NN \to NN\gamma$ bremsstrahlung contains
the excitation of $\Delta^{+,0}$ and the subsequent decay
$\Delta \to N \gamma$ according to the diagrams in Figs.~1a - d.
The isospin $\frac32$ of the $\Delta$ requires the coupling
to the isovectors
$\bf \pi$ and $\bf \rho$ in the corresponding vertices.
The parameters of the $\Delta N \pi$ and $\Delta N \rho$
vertices are adjusted by fitting both the
calculated $\Delta$ production cross section and the $\Delta\to N\pi$ angular
distribution to experimental data. Unfortunately, these data
do not allow for a unique set of parameters for both
$\Delta N\pi$ and $\Delta N\rho$ vertices. One of the possible sets
even does not need a $\rho$ exchange  contribution \cite{dileptons1}.
This is consistent with the parametrization of Ref.~\cite{Mosel1},
where the relative amplitude of the $\rho$ exchange is
about one order of magnitude smaller than that for the $\pi$ exchange, i.e.,
\begin{eqnarray}
\frac{T_\rho}{T_\pi}
\sim
\frac{f_{N\Delta\rho}}{f_{N\Delta\pi}}
\frac{m_\pi}{m_\rho}
\frac{\sqrt{g^2_{NN\rho}}}{\sqrt{g^2_{NN\pi}}}
\simeq 0.1.
\end{eqnarray}
Moreover, this ratio $T_\rho/T_\pi$
decreases if form factors with cut-off parameters
are included. Relying on this fact and trying
to use in our calculations as few as possible parameters,
we restrict ourselves in this subsection to resonance excitation via
$\pi$ exchange alone.

The $NN \to N \Delta$ interaction is characterized by the amplitude
\begin{eqnarray}
{\cal T}^\Delta =
g_\Delta\,F_\Delta^2 (q^2)
(\bar\psi^\Delta_\mu \bbox{T} q^\mu \psi)
(\bar\psi \bbox{\tau} \gamma_5 
\psi) \, (m_\pi^2 - q^2)^{-1},
\label{d1}
\end{eqnarray}
where the form factor $F_\Delta$ has the same form as in Eq.~(\ref{eq7})
but now with the cut-off $\Lambda_\Delta$. The isospin transition
operator ${\bf T}$ is defined as in Ref.~\cite{Brown}.
The propagator of the $\Delta$ is used in the standard form
for Rarita - Schwinger fields
\begin{eqnarray}
S^{\mu \nu}_\Delta (p) = \frac{-(\gamma^\alpha p_\alpha + M_\Delta)}
{(p^2 - M_\Delta^2 + i M_\Delta \, \Gamma_\Delta(p^2))}
[g^{\mu \nu} - \frac13 \gamma^\mu \gamma^\nu -
\frac23 M_\Delta^{-2} p^\mu p^\nu -
\frac13 M_\Delta^{-1} (\gamma^\mu p^\nu - p^\mu \gamma^\nu) ],
\end{eqnarray}
with $M_\Delta =$ 1.232 GeV and
momentum dependent width $\Gamma_\Delta(p^2)$ as
in Ref.~\cite{delta_width}
\begin{eqnarray}
\Gamma_\Delta(p^2)
& = &
\Gamma_\Delta^0\left[\frac{k(p^2)}{k(M_\Delta^2)}\right]^3
\frac{k^2(M_\Delta^2)+c^2}{k^2(p^2)+c^2},
\quad
c = 200 \mbox{MeV}, \\
\nonumber\\
\nonumber\\
k(p^2)& = &
\sqrt{\frac{(p^2+M_N^2-m_\pi^2)^2}{4p^2}-M_N^2}.
\label{eq16}
\end{eqnarray}

The effective strength parameter
$g_\Delta$ in Eq.~(\ref{d1}) is normalized to the
VerWest - Arndt parametrization of the total
$\Delta$ production cross section $\sigma_\Delta^{WA}$~\cite{West-Arndt}
\begin{eqnarray}
g_\Delta^2 = \frac
{\sigma_\Delta^{WA}}
{\int dt \,(d \sigma_\Delta/dt)},
\end{eqnarray}
where the angular differential cross section
\begin{equation}
\frac{d\sigma_\Delta}{dt} =
\frac{1}{16\pi{s(s-4M_N^2)}}\,|\bar{ {\cal T}^\Delta}|^2
\label{eq18}
\end{equation}
is calculated within our model with direct and exchange terms.
The cut-off parameter $\Lambda_\Delta =$ 0.7 GeV
is fixed by the requirement to get a optimum reproduction
of the experimental data on the $\Delta$ production angular
distribution \cite{delta_data}, see Fig.~1 in Ref.~\cite{dileptons1}.

Finally, the decay vertex $\Delta \to N \gamma$ is parametrized by
\begin{eqnarray}
\Gamma_{\mu\alpha} = - i e G_\Delta (
\gamma^\nu k_\nu g_{\mu\alpha} -
\gamma_\alpha k_\mu) \gamma_5
\label{eq19}
\end{eqnarray}
with $G_\Delta =$ 2.18, which results in the decay width
$\Gamma^{\Delta \to N \gamma} =
\tilde B^{\Delta \to N\gamma} \Gamma_\Delta^0$, where
$\tilde B^{\Delta \to N \gamma} = 6 \cdot 10^{-2}$
is the branching ratio, and
$\Gamma_\Delta^0 =$ 0.12 GeV denotes the total $\Delta$ width.


\section{Results of bremsstrahlung calculations}

Now we are going to present the results of our numerical calculations of the
above cross sections.
Let us first consider the contribution of the anomalous magnetic
moment of the nucleon in the vertex Eq.~(\ref{eq10}) at $T_{\rm kin}$=1 GeV.
For the sake of simplicity we show in Fig.~3
the triple differential cross section
${\omega d^3 \sigma}/{d \omega \, d\Omega_\gamma}$
at fixed angle of the outgoing photon.
The contribution of the anomalous moment is
proportional to $\omega^4$ and becomes important at relatively large
photon energies, i.e.,  at $\omega > \omega_{\rm max}/2$,
where $\omega_{\rm max}={(s - 4 M_N^2)}/2\sqrt{s}$.
At low photon energies only the electric
part of the electromagnetic current contributes,
and the cross section is well described
by the conventional soft-photon approximation.
The relative effect of the anomalous magnetic moment
in $pp$ scattering is larger than in $pn$ interactions.

The magnetic structure of the vertex in Eq.~(\ref{eq11})
leads to a similar behavior of the contribution from the
internal $V \to \pi \gamma$ transition, but its absolute value
is much smaller in the 1 GeV region as shown in Fig.~4.
This figure is more of methodical interest because at initial
energy above the $\Delta$ production threshold and at
$\omega >$  0.2 - 0.4 $\omega_{\rm max}$, the
contribution of the $\Delta$ channel becomes anyhow dominant.
Below the $\Delta$ threshold the $\omega/\rho \to \pi \gamma$ emission
has the same order of magnitude as the magnetic terms in the non-resonance
bremsstrahlung.

We have checked that the internal radiative meson conversion
and the anomalous magnetic moment of the nucleons do not change,
on a visible level, our previous results \cite{PLB} on the differential
cross sections $d \sigma /d \omega$ and $d \sigma / d\Omega_\gamma$.
Therefore we are going to discuss the cross sections for more specific
geometries which are easily accessible at COSY-TOF. In particular, we
select a non-coplanar geometry.
Our results of the differential cross sections
${\omega d^5\sigma}/{d \omega \,d\Omega_\gamma d\Omega_{p'}}$
are displayed in Figs.~5 and 6 for $pn$ and $pp$ reactions at beam energies of
$T_{\rm kin} =$ 0.7, 1.0, 1.35 and 1.7 GeV and fixed angles in the cms.
This energy interval covers the
expected range of validity of our model. With increasing beam energy the
r\^ole of $\Delta$ excitations becomes more important.

Let us first consider the reaction $pn \to pn \gamma$ in more detail.
In Fig.~5a one can distinguish two regions
in the energy distribution of the differential cross section.
At low photon energies the pure bremsstrahlung dominates,
while above $\omega/\omega_{\rm max} =$ 0.4 - 0.5
most photons stem from the $\Delta$ decays.
Interference effects are not important in these spectra
when considering the total yield.
The situation, however, changes when considering the
angular distribution at $T_{\rm kin} =$ 700 MeV
for $\omega/\omega_{\rm max} =$ 0.5 (Fig.~5b).
Here the interference terms become as large
as the dominating individual contributions at large angles.
(Note that in Ref.~\cite{Mosel1} the strongest interference
effects are also found near the kinematical boundary.)
Since we keep
$\omega/\omega_{\rm max} \approx$ 0.5,
this effect is not seen for higher bombarding
energies because the value of $\omega/\omega_{\rm max}$,
where the bremsstrahlung and
$\Delta$ contributions are equally strong, is changed to smaller numbers.

Similar conclusions hold for the reaction $pp \to pp\gamma$ (see Fig.~6).
Here the cross over
in Fig.~6a from the bremsstrahlung to the $\Delta$ decay dominated region
is sharper. Interference terms are smaller than the dominating contributions.
The most remarkable fact is the double-humped bremsstrahlung spectrum
in the angular distribution in Fig.~6b,
which however is hidden under the $\Delta$ decay contribution.
The relative sign of the $\Delta$ decay - bremsstrahlung interference still
remains unknown. This problem was also explored in Ref.~\cite{Mosel1} for
di-lepton production spectra. As illustrated in Fig.~6b, the sign of
$\Delta$ decay - bremsstrahlung interference may be very important
at certain photon angels at higher beam energies.
In our consideration the interference term is positive at small photon angles
and low beam energies,
but at large photon angle and $T_{\rm kin}=$ 1.7 GeV
its sign changes and becomes negative.
The total angular distributions in Figs.~5 and 6
do not show drastic variations,
and it seems difficult to disentangle the different photon sources
with these observables.

An interesting observation comes from the ratio of the photon emission
cross sections for $pn$ to $pp$ reactions displayed in Fig.~7.
From Figs.~4 and 5 alone one can hardly
make a distinction between $pn$ and $pp$ bremsstrahlung.
They have similar shapes with respect to the photon energy distribution.
But one can see from Figs.~7a and b that these reactions may show quite
substantial (and measurable) differences.
As known, the ratio of the di-electron
cross sections for $pd$ to $pp$ reactions
up to 1.5 GeV has a non-monotonic behavior \cite{DLS}. The Figs.~7a and b
show that the ratio of bremsstrahlung cross sections $pn/pp$
has an analogous shape. At large photon energy the bremsstrahlung
$pn/pp$ ratio increases similar as the di-electron ratio $pd/pp$ does,
see Fig. 7a. It is also interesting to look on the photon angular distribution
in Fig. 7b. Here we predict a pronounced non-monotonic behavior of this
observable at $T_{\rm kin} =$ 1.3 - 1.7 GeV which stems from the
large contribution of the negative $\Delta$ decay - bremsstrahlung
interference term (cf. Fig.~6b).

Our results of the differential cross section
$\omega d\sigma/d\omega d\Omega_\gamma d \Omega_{p'}$
might be compared with the existing experimental data
on $pp \to pp \gamma$ at $T_{\rm kin} =$ 730 MeV \cite{Nefkens}.
One observes in Fig.~8 a good agreement of our model
(without efficiency corrections)
with the raw data \cite{Nefkens}
for the counters G07 and G10. At very small photon energies one recovers
the results of the soft photon approximation.
At $\omega > 100$ MeV the $\Delta$ contribution becomes dominant
which causes an increase of the cross section, in fair agreement with
the data. This has been already observed in Ref.~\cite{Jetter}.


\section{The $\eta \to \gamma \gamma$ contribution}

As mentioned in the Introduction the $\eta$
meson production and its radiative two-photon decay
plays an essential r\^ole in the di-electron
spectra at 1 GeV beam energy and above \cite{dileptons1}.
In this section we estimate the
$\eta$ contributions to the inclusive $NN\to X\gamma$ reaction by
integrating over the momentum of the second photon. This consideration has
rather methodical character, because if one looks at the inclusive
one-photon distribution from intermediate
pseudo-scalar meson decay into two photons,
one should also consider the process
$\pi^0 \to \gamma\gamma$ as well, which is expected to give
a rather large contribution because of the large pion production cross section.
In the di-electron spectra at 1 GeV beam energy this problem does not appear
since the contributions of
$\pi^0 \to \gamma^*\gamma$ and $\eta\to \gamma^*\gamma$
(where $\gamma^*$ is virtual photon with invariant mass $M^2>0$)
are separated, because the $\pi^0$ decay
contributes only at small invariant mass $M^2<m_\pi^2$.
But the $\eta$ decay contribution may become important and
may be as large as the dominant $\Delta$ decay contribution
or even larger \cite{Mosel3}. Here, we intent to study
the $\eta$ decay contribution in comparison with the $\Delta$ channel
and the interference contribution from $\Delta$ decay and non-resonance
$NN$ bremsstrahlung  as an additional independent test
of the strong effect of the $\eta$ Dalitz decay channel near the threshold.
In doing so we assume that photons from $\pi^0$ decays
need not to be considered,
since they can be identified and rejected experimentally \cite{Nefkens}.

Relying on the fact that the $\eta\to \gamma\gamma$ decay width
is rather small, we get the following expression
for the $\eta$ contribution to the triple differential cross section
\begin{eqnarray}
\frac{d \sigma}{d \omega \, d\Omega}
=\frac{\omega \tilde B^{\eta \to \gamma \gamma} \, \sigma_\eta(s)}
{8 \, (2\pi)^6 \sqrt{s \, (s - 4 M_N^2)} }
\left[\frac{ \int d\Omega_{p'} \, d\varphi_\eta
\int\limits_{E_{\rm min}(\omega)}^{E_{\rm max}}
|\overline{ {\cal M}_{\eta} }|^2
D(p') \, dE_\eta}
{\int d\Omega_{p'} \, d\Omega_\eta
\int\limits_{m_\eta}^{E_{\rm max}}
|\overline{ {\cal M}_{\eta} }|^2
D(p') \, p_\eta \, dE_\eta }\right],
\label{eta1}
\end{eqnarray}
where $\sigma_\eta (s)$ is the $\eta$ production cross section;
$E_{\rm min,\rm max}$ are the minimum and maximum values of the
$\eta$ meson energy which depend on the photon energy via
\begin{eqnarray}
E_{\rm max} =(s + m_\eta^2 - 4 M_N^2)/(2 \sqrt{s}),\,\,\,
E_{\rm min}(\omega) = \omega + m_\eta^2/(4 \omega), \nonumber
\end{eqnarray}
and
$D(p')$ and $E'$ are defined as in Eq.~(\ref{eq3}), but now with
\begin{equation}
A = 2 (\sqrt{s} - E_\eta), \,\,\,
B = s - 2 \sqrt{s} E_\eta + m_\eta^2, \,\,\,
C = 2 p_\eta \cos \Theta_{\vec p_\eta \vec p'}.
\nonumber
\end{equation}
We use the known branching
ratio $\tilde B^{\eta \to \gamma \gamma} =$ 0.39.
When considering the amplitudes for the process $NN \to NN \eta$ we assume
that the excitation and the decay of the $N_{1535}$ resonance
dominates the $\eta$ production. We calculate the
$N_{1535}$ excitation within the OBE model with
$\pi^{\pm,0}, \rho^{\pm,0}, \omega, \sigma$ exchange which contribute
to the matrix elements according to the post-emission diagrams in
Figs.~1g and h
$$^{NN}{\cal M}_\eta =
\sum_{M=\pi,\rho,\omega,\sigma}
{^{NN}\hat {\cal M}^{(g,h)}_M},$$
where for $pp$ reactions only the exchange of neutral mesons
needs to be considered,
whereas for the $pn$ reaction also the exchange of charged mesons is included.
Note that the pre-emission contributions are considerably smaller and can be
neglected.

The matrix elements are constructed according to the
interaction Lagrangian
\begin{eqnarray}
{\cal L}_{int}' =
-i g^*_\pi \bar\psi_* \bbox{\tau} \psi \bbox{\pi} +
g^*_\rho \bar\psi_* \gamma_5 \gamma_\mu \bbox{\tau} \psi
\bbox{\rho}^\mu +
g^*_\omega \bar\psi_* \gamma_5 \gamma_\mu \psi \omega^\mu +
g^*_\sigma \bar\psi_* \gamma_5 \psi \sigma +
g^*_\eta \bar\psi_* \psi \eta,
\end{eqnarray}
where $\psi_*$ denotes the $N_{1535}$ wave function and
$g^*_M=g^{*0}_M \,F_M$ are the corresponding coupling constants
with the same form factors according to Eq.~(\ref{eq7}) as in $NNM$ vertices.
Evaluating the $\eta$ production cross section
we use same procedure as in the $\Delta$ production, i.e., the
normalization of $\sigma_\eta(s)$ to the known experimental value.
In our calculations we employ the parametrization
\begin{eqnarray}
\sigma_\eta = \frac{a \, (1 - \xi^2)}{ (1 + [b (\xi^{-1} - 1)]^c)
\sqrt{s (s - 4 M_N^2)} },
\label{eta_cs}
\end{eqnarray}
with
$a = 4 \cdot 10^2$ ($3 \cdot 10^3$) mb$\cdot$GeV$^2$,
$b =$ 17 (33) and $c =$ 1.8 (2.1) for
$pp$ ($pn$) reactions. The quantity $\xi$ is defined as
$\xi = \sqrt{s_0/s}$ with $s_0 = (2 M_N + m_\eta)^2$.
This parametrization is in agreement with results of
Ref.~\cite{scaling} and numerically coincides with
Ref.~\cite{eta} and has been used
for di-electron production \cite{dileptons1}.
Eq.~(\ref{eta_cs}) allows to constrain one parameter
in the set of coupling constants, and as a result we
need only the ratios of
$g^{*0}_\pi : g^{*0}_\rho : g^{*0}_\omega : g^{*0}_\sigma =
1 : 0.78 : 0.55 : 0.17$ which are taken from Ref.~\cite{scaling}.

Eq.~(\ref{eta1}) shows that the $\eta$ decay contributes in the window
$\tilde\omega_{\rm min} \le \omega \le \tilde\omega_{\rm max}$ with
$\tilde\omega_{\stackrel{\rm max}{\rm min}} =
\frac{1}{2} (E_{\rm max} \pm \sqrt{E_{\rm max}^2 - m_\eta^2})$.
The $N_{1535}$ resonance is assumed to be an unstable particle, which can
be described by replacing $M_N \to M^* - i \frac{1}{2} \Gamma^*$
in the nucleon propagator with $M^* = 1535$ MeV and
$\Gamma^* =$ 180 MeV.

Our results are displayed in Figs.~9a and b for $pn$ and $pp$ collisions
at $T_{\rm kin} =$1.35 and 1.7 GeV.
The energy $T_{\rm kin} =$ 1.35 GeV
is just slightly above the $\eta$ threshold.
At $T_{\rm kin} =$ 1.7 GeV there is a narrow window wherein the
$\eta$ decay contribution almost shines out. Within this window
the $\eta$ channel is as large as the dominant $\Delta$ decay contribution,
which is in agreement with previous conclusions \cite{dileptons}.
However, at extremely large photon energies $\omega \to \omega_{\rm max}$
the interference of the $\Delta$ channel with the pure bremsstrahlung process
becomes comparable with the dominant channel. That is again
in agreement with previous findings \cite{Mosel1}.
So we conclude that for a clear understanding of the inclusive
real and virtual photon spectra both $\eta$ decay and strong
interference of the resonance and non-resonance bremsstrahlung must be
taken into account especially for photon energies close
to the kinematical limit.


\section{$NN$ bremsstrahlung at $T_{\rm kin} =$ 280 MeV}

As mentioned in the Introduction, there is a large body of experimental data
on $NN$ bremsstrahlung
at beam energies
$T_{\rm kin}\sim$ 200 - 300 MeV around the pion production threshold,
and new precision data are to be expected soon
from COSY-TOF and other installations.
Therefore, it seems interesting to check a possible applicability
of the OBE model by comparing with this data.
One should stress, however, that the more elaborate potential models
are usually considered as adequate in this energy region,
since initial and final state interaction or rescattering
effects may be essential.
However, there is no standard potential model.
Various groups suggest their own models with
specific approximations. That makes a comparison between different approaches
rather difficult.
We mention in this context also a recent successful application of
the effective OBE model
to a description of meson production in a
wide energy interval down to the pion threshold \cite{Moselpi}.
All this inspires us to apply
our effective OBE model to the low energy bremsstrahlung data.
Note that it is not obvious that the effective two-body $T$ matrix
with parameters adjusted at much higher energies is adequate for
such low energy
$NN$ scattering. Nevertheless, we expect that the OBE model reproduces the
cross features of bremsstrahlung in this energy region.

We calculate the differential cross section
$d^5 \sigma/d\Omega_{p'_1} d\Omega_{p'_2}d\vartheta_\gamma$
for bremsstrahlung as a function
of the photon polar angle $\vartheta_\gamma$ at fixed
polar proton angles $\vartheta_{p_{1',2'}}$
at $T_{\rm kin} =$ 280 MeV.
A comparison of our results with the data \cite{Michaelian}
are shown in Fig.~10.
One can see that the OBE model describes, without any fine tuning of the
parameters, the structure of the
data quite satisfactory. Notice, however, that there is some debate on the
absolute normalization of the data \cite{V_Brown}.
The soft photon approximation does not so well account for the data.
This means that for qualitative analyses of unpolarized
bremsstrahlung observables one can use the effective OBE
model in a wide energy region from 2 GeV down to the pion threshold.


\section{Discussion and summary}

The present study of bremsstrahlung in $NN$ reactions relies on the
effective OBE model with parameters adjusted to elastic $NN$ scattering.
While a generalization of the OBE model to inelastic processes seems
straightforward by calculating a considerable number of Feynman
diagrams, one also meets not yet satisfactorily clarified items.
One such challenging issue concerns the effective strong and electromagnetic
form factors.

As mentioned in Section 2 we follow here Ref.\cite{Mosel1} and others
and neglect the half-off-shell effects in the strong $NNM$ form factors.
Since we consider the emission of real photons with $k^2 = 0$ the
electromagnetic form factor can still depend on the momentum
squared of the virtual nucleon. Desirable would be a unique procedure to
introduce off-shell effects in a consistent way for both the strong and the
electromagnetic form factors in the general case when the photon has
arbitary virtiality, i.e., $k^2 \ne 0$ as in di-electron production and
electron scattering. Various aspects of this problem have been studied
in Refs.~\cite{dileptons,Mosel1,Mosel,Naus,Doenges,Nyman,others}.
A outcome of some models \cite{dileptons,Mosel1}
seems to be the additional suppression of
form factors by off-shell effects, while other models \cite{Nyman}
point to increased cross sections by off-shell effects.
It is worth mentioning that probably it is not enough to put off-shell
form  factors into standard on-shell vertex functions.
Generally, new operator structures can appear \cite{Naus,Nyman}
and modify the features of the cross section.
Also the off-shell effects must be consistently considered
with the assumed $NN$ interaction. In our approach we take the OBE model
literally and use it without further modifications as proven successful
for virtual photon \cite{Mosel1} and pion production \cite{Moselpi}.
We feel that precision data are needed for getting a guideline for
improvements of the effective OBE model, e.g. by including loop contributions.

In summary we report a study of photon production within an effective
one-boson exchange model which is indented for a prediction
of forthcoming data at COSY-TOF. The present results focus on the
general theoretical scheme but compare also successfully with existing data.
We mention that accurate data on the
elastic scattering and the $\Delta$ contributions are needed to
get more confidence in the parameters to be employed.
It should be stressed that also the reaction
$pd \to pd \gamma, pn \gamma$ is very worth measuring
since it provides valuable information on the $pn$ channel.

{\small
{\bf Acknowledgments:}
Stimulating discussions with
E.L. Bratkovskaya, W. Cassing, H. Freies\-le\-ben, E. Grosse,
B. \& L. Naumann, K. M\"oller, and U. Mosel are gratefully acknowledged.
The work is supported by BMBF grant 06DR829/1 and
Heisenberg-Landau foundation.


\newpage
\begin{figure}
 \vskip 4.cm
 \centerline{\epsfxsize=0.9 \hsize \epsffile{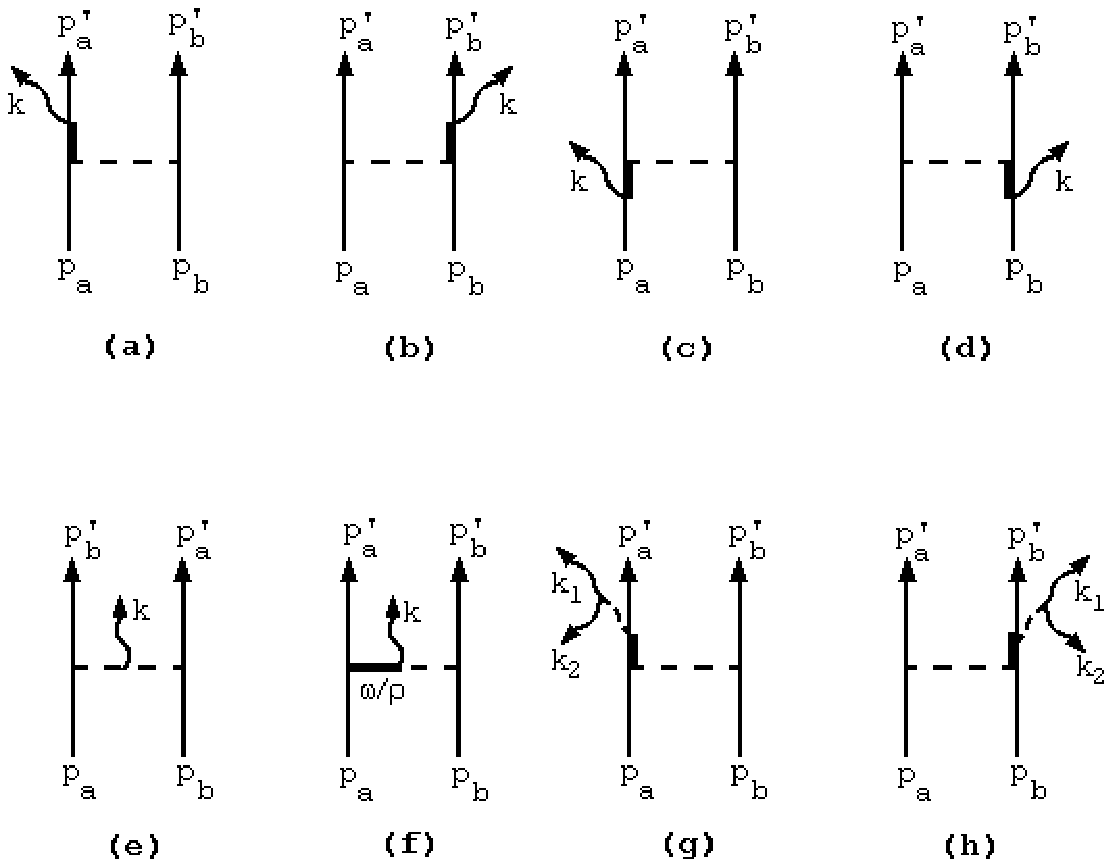}}
 \vskip 1cm
\caption{
The Feynman diagrams for $NN$ bremsstrahlung considered
in this paper.
(a) - (d): bremsstrahlung with an intermediate off-shell nucleon
or a $\Delta$
(fat lines);
(e): photon emission from the exchange of a charged meson,
(f): emission from the $V \to \pi\gamma$ transition, i.e.,
internal radiative meson conversion,
(g) and (h): the intermediate nucleon line represents a $N_{1535}$ state
which decays into a nucleon and an $\eta$, which undergoes subsequent
two-photon decay. The exchange digrams are not displayed.
}
\label{elas}
\end{figure}

\newpage

\hspace*{2.cm}

\begin{figure}
\vskip 1.cm
\centerline{\epsfxsize=0.7 \hsize \epsffile{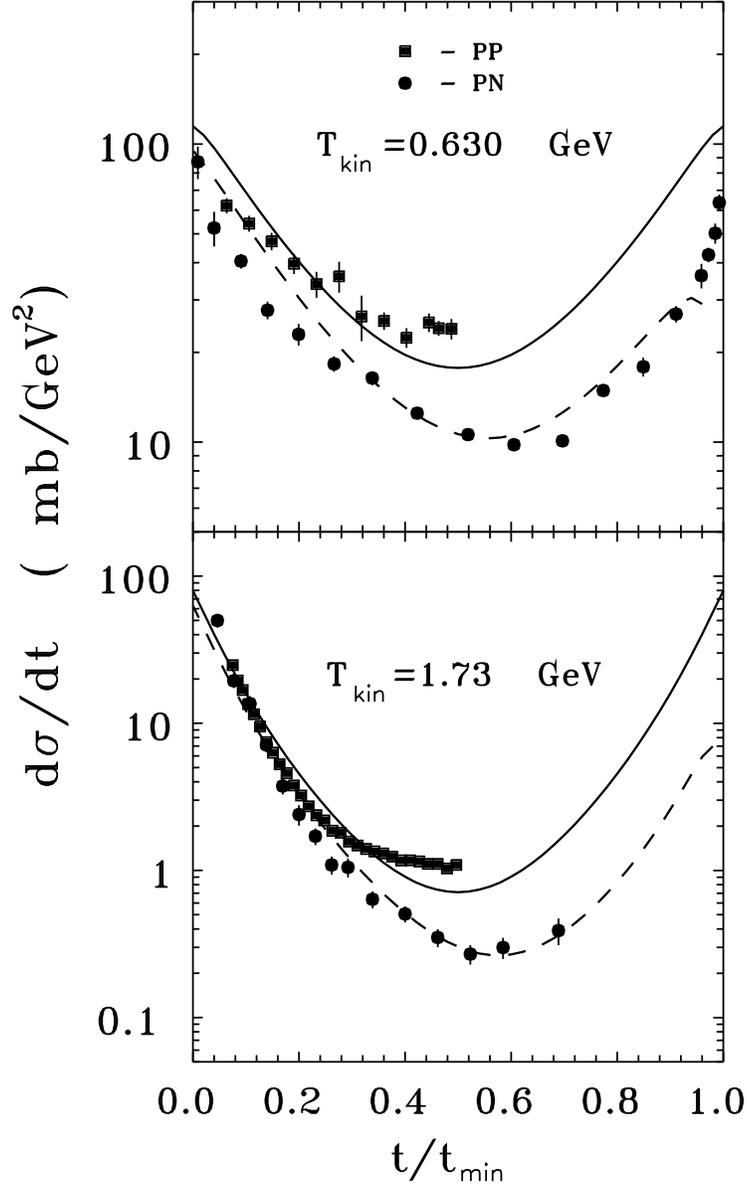}}
\vskip 1cm
\caption{
The elastic nucleon - nucleon cross section at
beam energies $T_{\rm kin} =$ 0.63 and 1.73 GeV. Solid and dashed lines
depict results for the $pp$ an $pn$ reactions.
Experimental data are taken from
Ref.~\protect\cite{cross_sections,elsc}.
$t_{\rm min} = 4 M_N^2 - s$.
}
\label{ell}
\end{figure}

\newpage

\hspace{-1.7cm}
\begin{figure}
 \vskip -.7cm
 \centerline{\epsfxsize=0.7 \hsize \epsffile{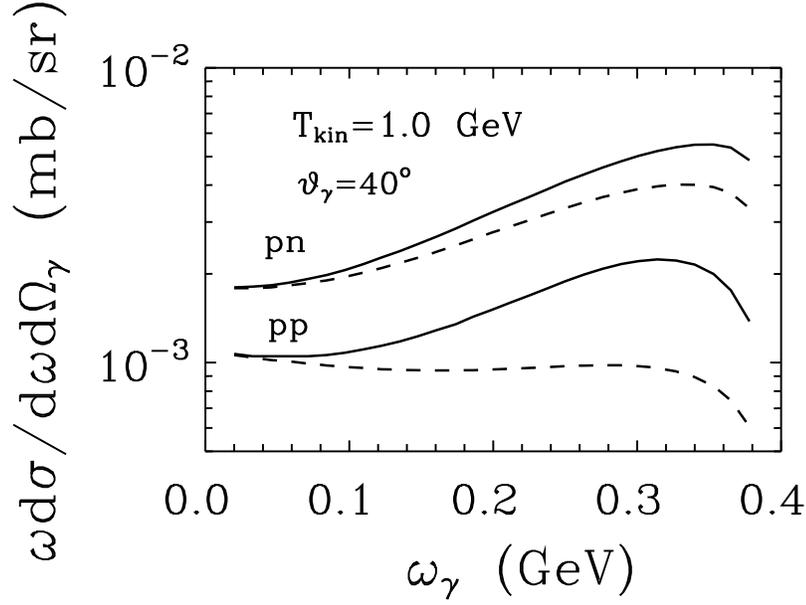}}
 \vskip -0.1cm
\caption{
The differential cross section  for proton - proton and
proton - neutron bremsstrahlung at $T_{\rm kin} =$ 1 GeV. The
cms photon angle is $\vartheta_\gamma=40^o$. The solid
and dashed lines correspond to calculations with
and without anomalous magnetic moment contributions.
}
\label{anml}
\end{figure}

\begin{figure}
 \vskip  .5cm
 \centerline{\epsfxsize=0.7 \hsize \epsffile{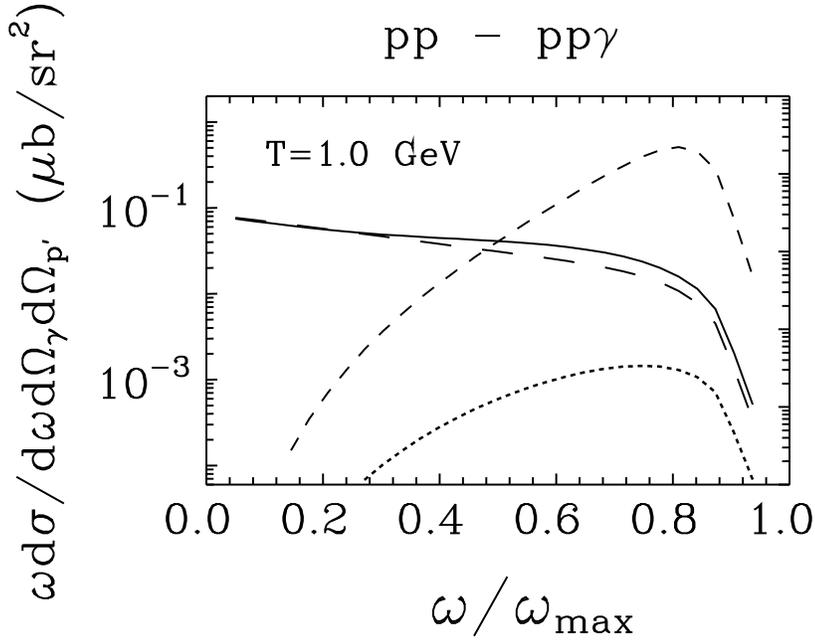}}
 \vskip .1cm
\caption{
The radiative meson decay $\omega,\rho \to \pi \gamma$
contribution to the bremsstrahlung
at $T_{\rm kin} =$ 1 GeV for fixed angles of the exit particles in cms:
$\vartheta_{p'}=40^o$,
$\vartheta_\gamma=60^o$, $\phi=60^o$.
The solid, short-dashed, long-dashed and dotted  curves correspond to
the contributions of non-resonance bremsstrahlung,
soft photon approximation, $\Delta$ decay and
$\omega, \rho$ decay, respectively.
}
\label{omg}
\end{figure}

\newpage
\begin{figure}
 \vskip 3.8cm
 \centerline{\epsfxsize=0.9 \hsize \epsffile{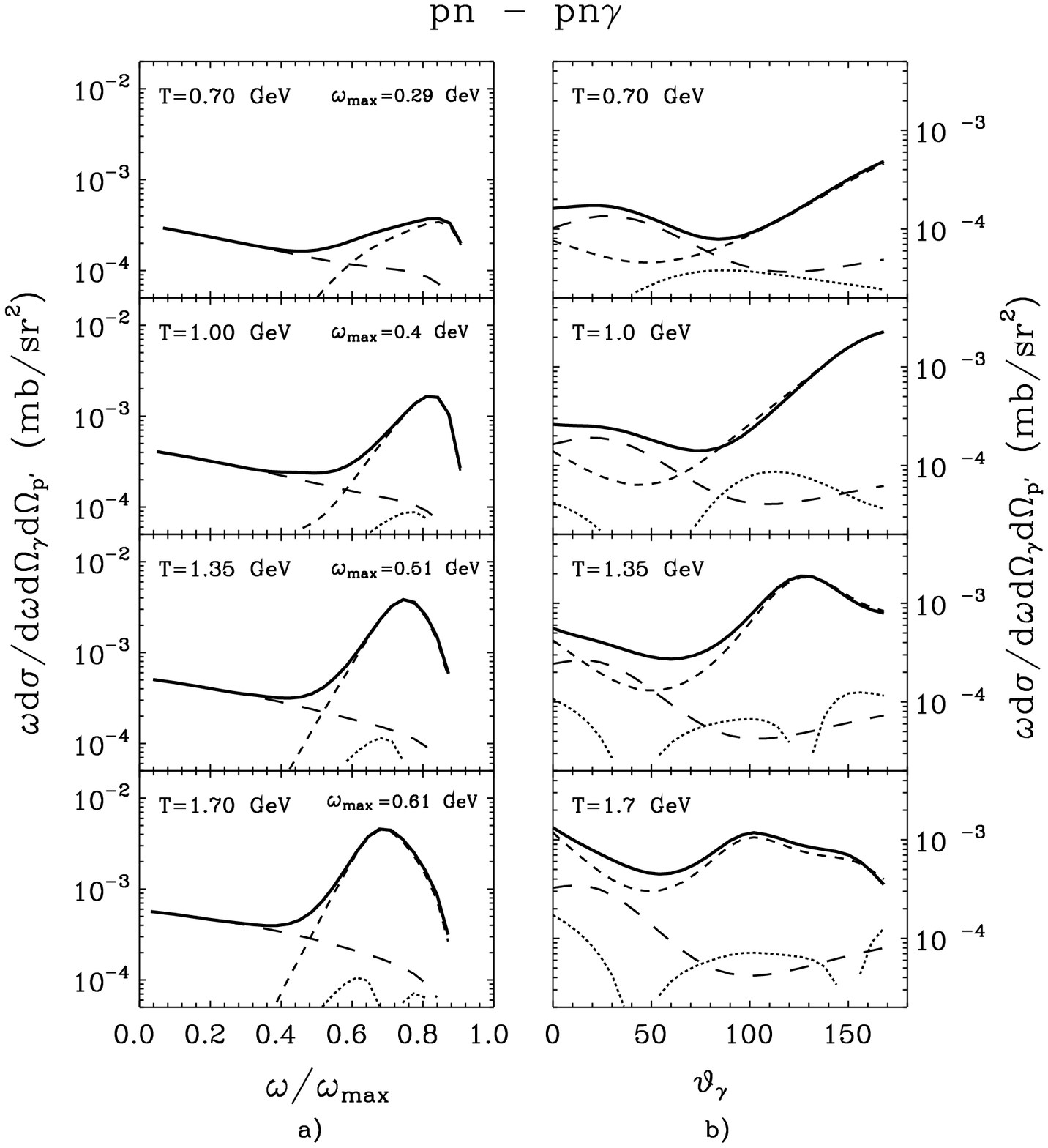}}
 \vskip 1cm
\caption{
The differential cross section
$\omega d\sigma/d\omega d\Omega_\gamma d\Omega_{p'}$ vs.
$\omega/\omega_{\rm max}$ in cms
at $\vartheta_{p'}=40^o$, $\vartheta_\gamma=30^o$, $\phi=60^o$) (a)
and vs. $\vartheta_\gamma$  at $\omega/\omega_{\rm max}=0.5$
at $\vartheta_{p'}=40^o$, $\phi=60^o$~ (b) for the reaction
$pn \to pn\gamma$ at $T_{\rm kin} =$ 0.7, 1.0, 1.35, 1.7 GeV
(the meaning of the various lines is the following one:
long dashed: bremsstrahlung, short dashed: $\Delta$ decay, dotted:
interference of bremsstrahlung and $\Delta$ decay, heavy solid lines:
total cross section).
}
\label{pn}
\end{figure}

\newpage
\begin{figure}
 \vskip 3.8cm
 \centerline{\epsfxsize=0.9 \hsize \epsffile{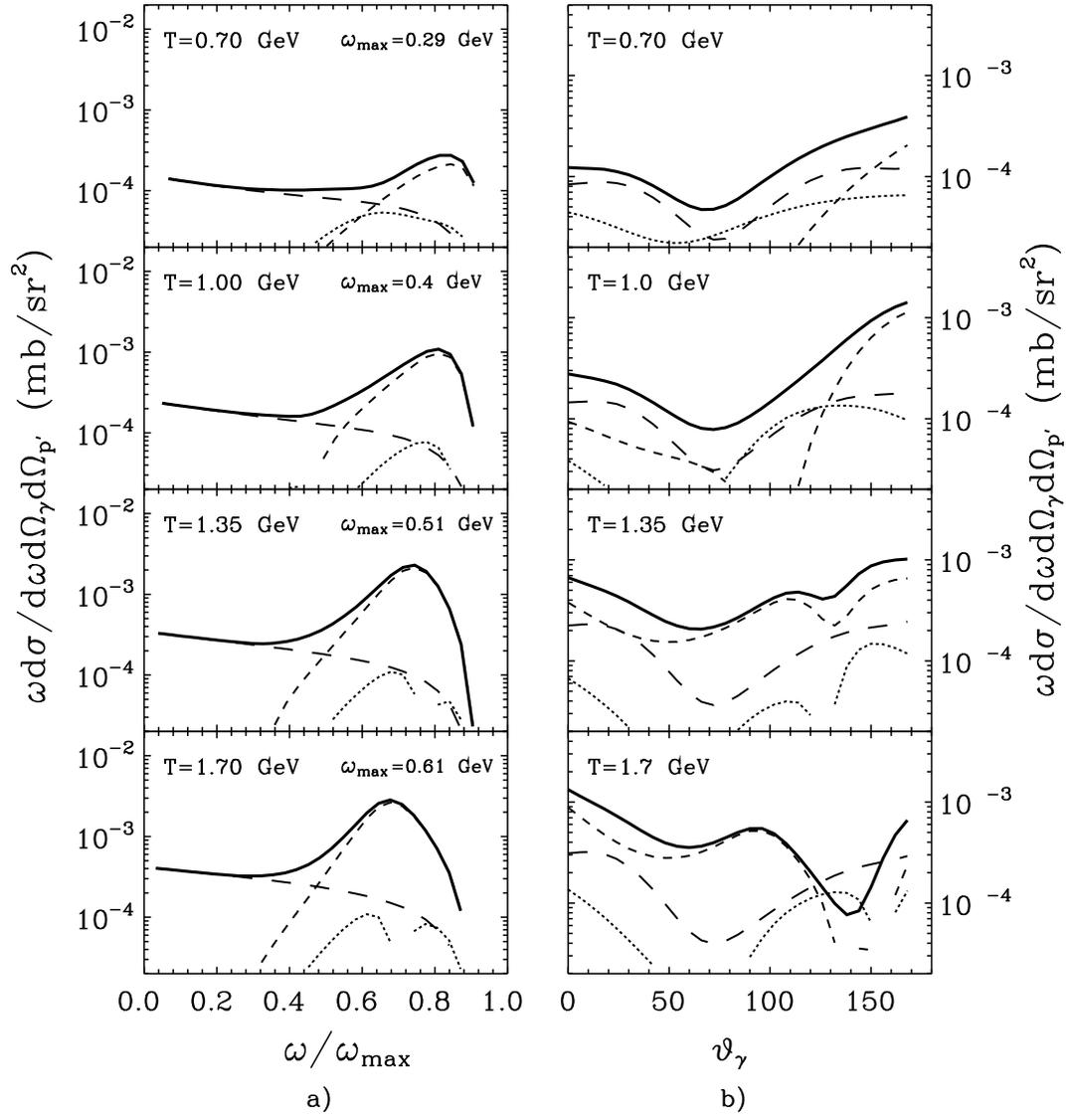}}
 \vskip 1cm
\caption{
The same as Fig.~\protect\ref{pn} but for $pp \to pp \gamma$.
}
\label{pp}
\end{figure}

\newpage
\begin{figure}
 \vskip 6cm
 \centerline{\epsfxsize=1. \hsize \epsffile{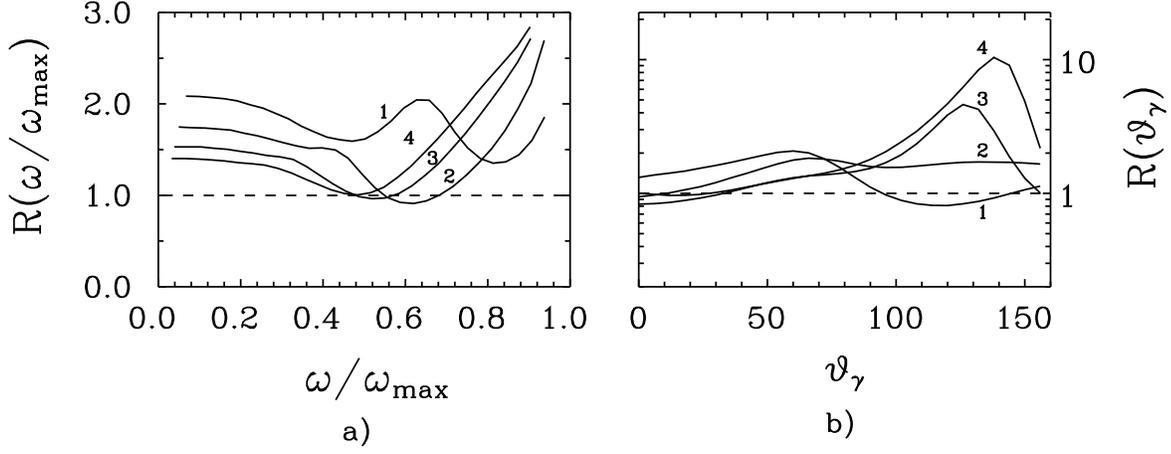}}
 \vskip 1cm
\caption{
The ratio
$(\omega d\sigma/d\omega d\Omega_\gamma d\Omega_{p'})^{pn}/
(\omega d\sigma/d\omega d\Omega_\gamma d\Omega_{p'})^{pp}$
of differential cross sections shown in Figs.~\protect\ref{pn} and
\protect\ref{pp} as a function of $\omega/\omega_{\rm max}$
at  $\vartheta_{p'}=40^o$, $\vartheta_\gamma=30^o$, $\phi=60^o$ (a),
and as a function of the photon angle
at $\vartheta_\gamma$ at $\vartheta_{p'}=40^o$, $\phi=60^o$,
$\omega/\omega_{\rm max}$=0.5 (b) for four energies
$T_{\rm kin}$=0.7, 1.0, 1.35, 1.7 GeV corresponding to the curves labeled
by 1 - 4.
}
\label{ratio}
\end{figure}

\begin{figure}
 \vskip 3.5cm
 \centerline{\epsfxsize=0.75 \hsize \epsffile{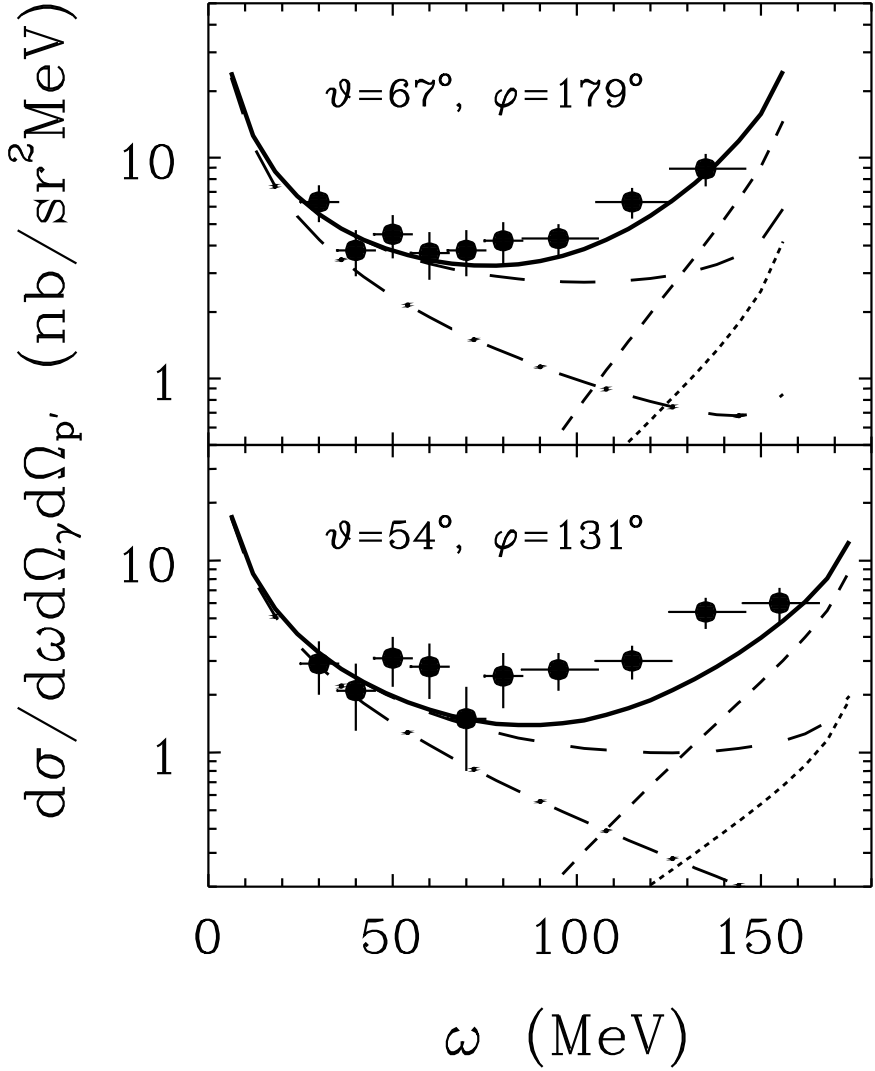}}
 \vskip 1cm
\caption{
Comparison of our calculations for the cross section of
the real photon production in proton - proton scattering at
$T_{\rm kin} =$ 730 MeV
with the experimental data for two geometries (counters G07 and G10).
The data are taken from the Ref.\protect\cite{Nefkens}.
Dot-dashed lines: soft photon approximation,
long-dashed lines: pure bremsstrahlung,
short-dashedlines: $\Delta$ decay,
dotted lines: $\Delta$ - bremsstrahlung interference.
The heavy solid curves represent the sum of these contributions.
}
\label{exp73}
\end{figure}

\newpage
\begin{figure}
 \vskip 2.cm
 \centerline{\epsfxsize=0.9 \hsize \epsffile{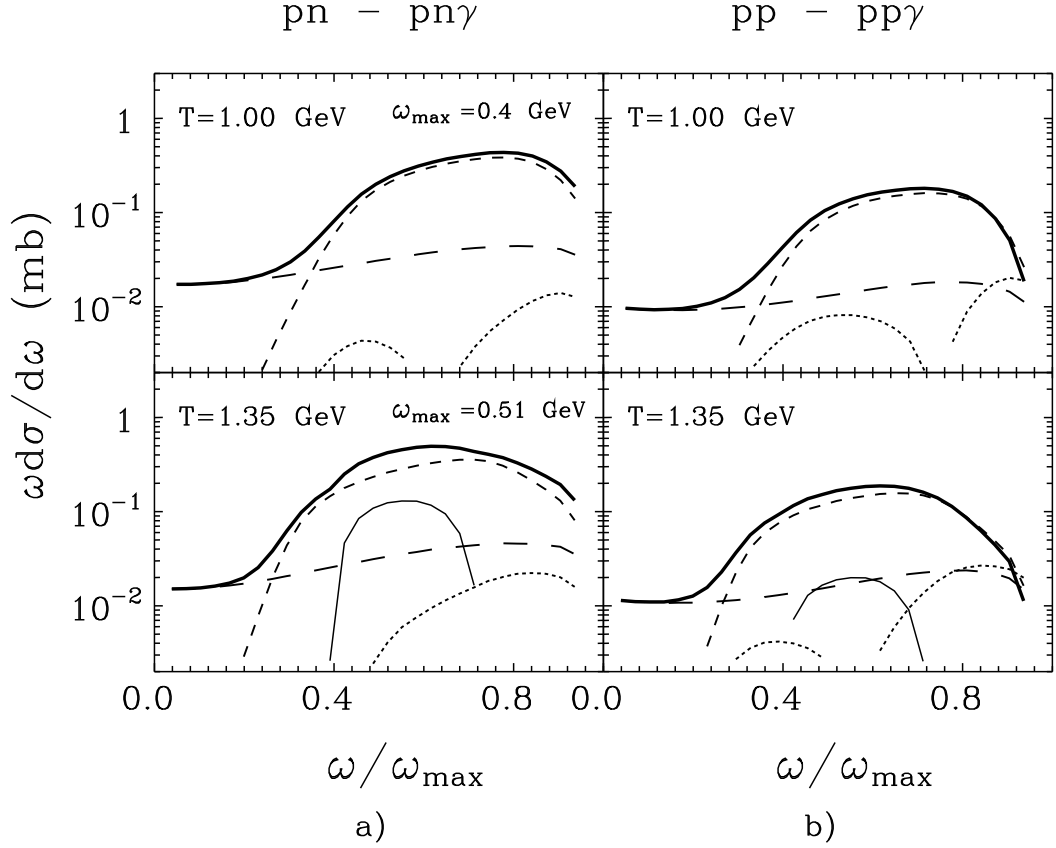}}
 \vskip 1cm
\caption{
The differential proton - neutron (a) and proton - proton (b)
cross sections in cms as a function of the photon energy for
illustrating the
$\eta$ decay contribution to the photon production in
comparison with others channels at two beam energies
$T_{\rm kin}=$ 1.0 and 1.35 GeV.
The long-dashed, short-dashed, dotted and solid lines represent
pure bremsstrahlung, $\Delta$ decay, $\Delta$ decay -
bremsstrahlung interference,
and $\eta$ decay. The heavy solids curve show the sum of these contributions.
}
\label{eta}
\end{figure}

\newpage
\begin{figure}
 \vskip 3.cm
 \centerline{\epsfxsize=.7 \hsize \epsffile{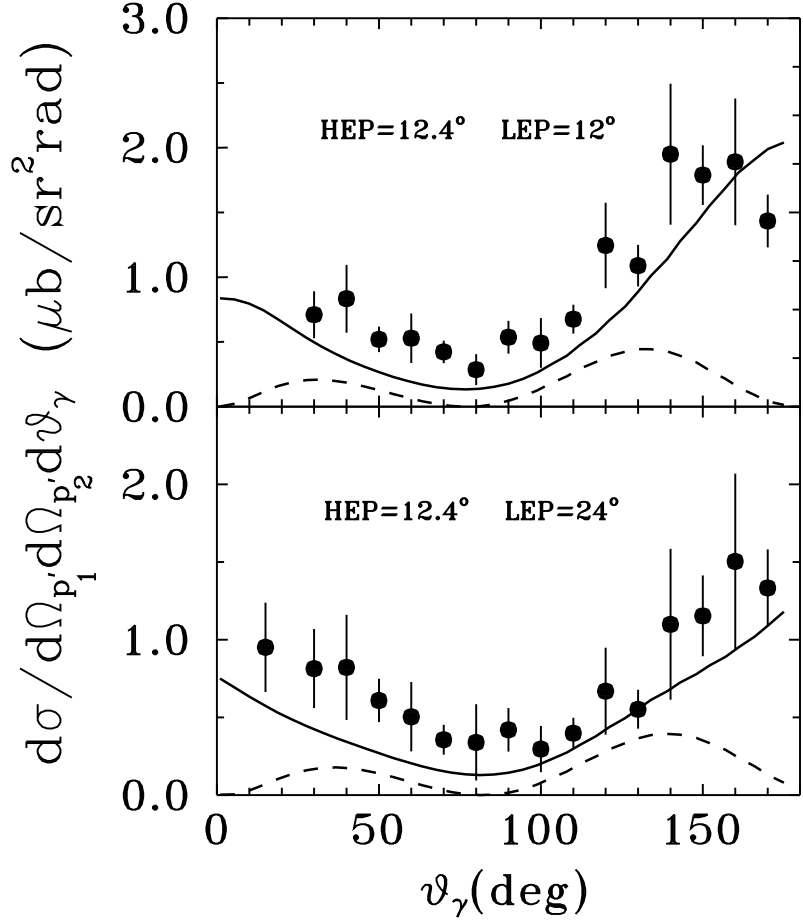}}
 \vskip 1cm
\caption{
Comparison of our calculations for the $pp \to pp \gamma$
cross section with experimental data at beam energy $T_{\rm kin} =$
280 MeV. The data are taken from
Ref.~\protect\cite{Michaelian} and scaled by a factor 2/3 as discussed
in Ref.~\protect\cite{V_Brown}.
The solid and dashed lines represent a pure bremsstrahlung calculation and
soft photon approximation.
The notion of HEP and LEP angles correspond to high-energy and
low-energy protons respectively.
}
\label{exp28}
\end{figure}

\newpage
\begin{table}
\caption{The parameter set used in the present work. \protect \\}
\begin{center}
\begin{tabular}{|ccrcc|}
\hline
meson    &     mass  & coupling & cut-off & energy parameter\\
M      & $m_M$ [GeV] & $g_M^0$      & $\Lambda_M$ [GeV] & $\beta_M$ [GeV$^{-1}$]\\
\hline
$\pi$    & 0.138      & 11.7        & 0.75 & 0.047 \\
$\rho$   & 0.770      & 1.78        & 1.60 & 0.047 \\
$\omega$ & 0.782      & 9.60        & 1.23 & 0.035 \\
$\sigma$ & 0.550      & 2.03        & 1.10 & 0.041 \\
\hline
\end{tabular}
\end{center}
\end{table}
\end{document}